\documentclass[11pt]{article}
\usepackage[utf8]{inputenc}
\usepackage[T1]{fontenc}

\usepackage[nocompress]{cite}
\usepackage{IEEEtrantools,stackengine}
\stackMath

\usepackage{authblk}
\usepackage{fullpage}
\usepackage{amssymb,amsmath}
\usepackage{float}
\usepackage{bm}
\usepackage{makecell}
\usepackage{siunitx}
\usepackage{csquotes}

\usepackage{xspace}
\newcommand\ie{i.e.\xspace}
\newcommand\eg{e.g.\xspace}

\DeclareMathOperator{\logit}{logit}

\def\sym#1{\ifmmode^{#1}\else\(^{#1}\)\fi}
\usepackage{booktabs}
\usepackage{longtable}
\usepackage{multirow}
\usepackage[export]{adjustbox}

\usepackage[dvipsnames]{xcolor}

\usepackage[left]{lineno}

\usepackage[
	colorinlistoftodos,
	textsize=footnotesize,
]{todonotes}

\definecolor{darkgreen}{rgb}{0.0, 0.5, 0.0}

\usepackage{setspace}
\usepackage[unicode=true]{hyperref}
\hypersetup{breaklinks=true,
	bookmarks=true,
	colorlinks=true,
	pdfborder={0 0 0},
	allcolors=blue}
\expandafter\def\expandafter\UrlBreaks\expandafter{\UrlBreaks
	\do\-}

\usepackage[%
	sort&compress
]{cleveref}
\Crefname{appendix}{Supplement}{Supplements}

\usepackage{times}
\usepackage{enumerate}
\usepackage{float}

\usepackage{subcaption}
\usepackage{graphicx}

\graphicspath{{./figures/}}

\usepackage[figuresright]{rotating}
\usepackage{tabularx}
\usepackage{dcolumn}
\usepackage{pdflscape}

\usepackage{array}

\newcolumntype{L}[1]{>{\raggedright\let\newline\\\arraybackslash\hspace{0pt}}p{#1}}
\newcolumntype{C}[1]{>{\centering\let\newline\\\arraybackslash\hspace{0pt}}p{#1}}
\newcolumntype{R}[1]{>{\raggedleft\let\newline\\\arraybackslash\hspace{0pt}}p{#1}}

\usepackage{sectsty}
\subsubsectionfont{\normalfont\itshape}

\makeatletter
\renewcommand{\fps@figure}{H}
\renewcommand{\fps@table}{H}
\makeatother

\allowdisplaybreaks

\usepackage{eurosym}

\usepackage{comment}

\usepackage{ntheorem}
\theoremseparator{:}

\usepackage[shortcuts]{extdash}
\usepackage{paralist}
\usepackage{colortbl} 
\usepackage{tabu}
\usepackage{bbm}
\usepackage{soul}

\DeclareSIUnit\week{week}
\DeclareSIUnit\weeks{weeks}
\DeclareSIUnit\billion{billion}

\begin{document}


\title{\centering\LARGE\singlespacing References to unbiased sources increase the helpfulness of community fact-checks}

\renewcommand\Affilfont{\fontsize{9}{10.8}\selectfont}

\author[1]{Kirill Solovev}
\author[1]{Nicolas Pröllochs\thanks{Correspondence: Nicolas Pröllochs (\url{nicolas.proellochs@wi.jlug.de})}}

\affil[1]{JLU Giessen, Germany}

\date{}

\maketitle


\begin{abstract}
\normalfont
\noindent
Community-based fact-checking is a promising approach to address misinformation on social media at scale. However, an understanding of what makes community-created fact-checks helpful to users is still in its infancy. In this paper, we analyze the determinants of the helpfulness of community-created fact-checks. For this purpose, we draw upon a unique dataset of real-world community-created fact-checks and helpfulness ratings from X's (formerly Twitter) Community Notes platform. Our empirical analysis implies that the key determinant of helpfulness in community-based fact-checking is whether users provide links to external sources to underpin their assertions. On average, the odds for community-created fact-checks to be perceived as helpful are 2.70 times higher if they provide links to external sources. Furthermore, we demonstrate that the helpfulness of community-created fact-checks varies depending on their level of political bias. Here, we find that community-created fact-checks linking to high-bias sources (of either political side) are perceived as significantly less helpful. This suggests that the rating mechanism on the Community Notes platform successfully penalizes one-sidedness and politically motivated reasoning. These findings have important implications for social media platforms, which can utilize our results to optimize their community-based fact-checking systems.
\end{abstract}

\flushbottom
\maketitle
\thispagestyle{empty}


\sloppy
\raggedbottom


\clearpage

\section*{Introduction}

Misinformation on social media can have real-world consequences. Among other instances, negative effects of misinformation have been repeatedly observed in the contexts of public safety \cite{Starbird.2017,Bar.2023,Jakubik.2023,Geissler.2023}, public health \cite{Broniatowski.2018,Gallotti.2020,Solovev.2022b}, and elections \cite{Aral.2019,Bakshy.2015,Grinberg.2019}. Recognizing this, professional fact-checkers and fact-checking organizations (\eg, \url{snopes.com}, \url{politifact.com}) routinely fact-check social media rumors in order to identify potentially misleading information on social media \cite{Vosoughi.2018}. However, due to restricted resources, these organizations struggle to keep up with the volume of content generation \cite{Micallef.2020}. Hence, recent research has advocated for delegating the fact-checking of social media posts to non-professional fact-checkers in the crowd \cite{Micallef.2020,Bhuiyan.2020,Pennycook.2019,Epstein.2020,Allen.2020,Allen.2021,Godel.2021}. A community-based approach to fact-checking is promising as it offers the capacity to conduct numerous fact-checks at high frequency and low costs \cite{Allen.2021,Pennycook.2019}. It may also address the trust issues observed with professional fact-checks \cite{Poynter.2019}. Recent experiments show that while the judgement of individual fact-checkers can be inconsistent and unreliable \cite{Woolley.2010}, even fairly small groups of non-experts can achieve an accuracy comparable to those of experts \cite{Bhuiyan.2020,Epstein.2020,Pennycook.2019,Godel.2021,Allen.2021}.

However, while the crowd may be \emph{capable} to accurately detect misinformation, it does not automatically entail that all users will \emph{decide} to do so \cite{Epstein.2020}. Crucial challenges encompass lack of engagement in critical thinking \cite{Pennycook.2019b}, politically motivated reasoning \cite{Kahan.2017}, and manipulation attempts \cite{Luca.2016}. Each of these behaviors can reduce the effectiveness of community-based fact-checking systems. For instance, there could be purposeful efforts by users to manipulate the fact-checking process by reporting social media contents as misleading based purely on non-conformance with their preconceptions or to achieve partisan ends \cite{Luca.2016}. Furthermore, the high level of (political) polarization among social media users \cite{Conover.2011,Barbera.2015} can lead to significantly different interpretations of facts or even entirely different sets of accepted facts \cite{Otala.2021}. Hence, crucial requirements in community-based fact-checking systems are sophisticated rating systems and fact-checking guidelines that promote helpful fact-checks. However, little is known regarding the question of what makes a community fact-check helpful.

Previous research has analyzed determinants of helpfulness in the context of customer reviews on online platforms such as \url{Amazon.com} and \url{Yelp.com}, yet not for community-created fact-checks of social media posts. For example, earlier works have found that meta-characteristics such as the age of the review, the rating, and the length are important determinants of the helpfulness of customer reviews \cite{Yin.2016,Mudambi.2010,Lutz.2022,Schlosser.2011,He.2015,Pan.2011}. Yet, despite apparent similarities, community-based fact-checking on social media substantially differs from customer reviews. While customer reviews commonly share (subjective) personal experiences with a product, the goal in fact-checking on social media is to carry out an accurate (and objective) assessment of a social media post. Furthermore, community-based fact-checking on social media must deal with politically biased views and a highly polarized user base. Ensuring high levels of trust with the fact-checkers' assessments is thus comparatively more important -- and more difficult to attain. To this end, modern community-based fact-checking systems typically step away from exclusively labeling potentially misleading social media content. Instead, they encourage users to write short textual fact-checking assessments and link to external sources (\eg, media outlets, scientific papers) to underpin their assertions. However, an understanding of how (and which) external sources affect the helpfulness of community-created fact-checks is largely missing. Shedding light on this question represents the goal of this research.

In the present work, we conduct an empirical analysis of the relationship between external sources in community-created fact-checks on social media and their perceived helpfulness. For this purpose, we utilize a unique dataset encompassing community-created fact-checks for social media posts obtained from X's ``Community Notes'' platform (formerly ``Birdwatch'') \cite{Twitter.2021}. In contrast to earlier (small-scale) crowd-based fact-checking initiatives, Community Notes allows users to identify misleading posts \emph{directly} on X. Specifically, the Community Notes feature allows users to tag posts they consider misleading and supplement them with written notes that provide context to the post (\eg, by referring to external sources). An integral feature of Community Notes is that it implements a rating system, which provides users with the capability to rate the helpfulness of notes contributed by other users. These ratings are intended to facilitate the identification of the context which people find most helpful. For our analysis, we gather all community-created fact-checks from the Community Notes during an observation period of more than 8 months. Subsequently, we implement explanatory regression models to holistically analyze how the presence of external sources is linked to the helpfulness of the corresponding fact-check. Furthermore, we study how the helpfulness varies with regards to the level of political bias of the external sources provided in the community-created fact-checks.

Our empirical analysis implies that linking to external sources is \emph{the} key determinant of helpfulness in community-based fact-checking. On average, the odds for community-created fact-checks to be perceived as helpful are 2.70 times higher if they link to external sources. Furthermore, we demonstrate that the helpfulness of community-created fact-checks varies depending on their level of political bias. Here, we find that community-created fact-checks linking to high bias sources (of either political side) are perceived as significantly less helpful. This suggests that the rating mechanism on the Community Notes platform successfully penalizes one-sidedness and politically motivated reasoning. Our findings have important implications for social media platforms that can utilize our results to optimize their fact-checking guidelines and promote helpful fact-checks.

\section*{Background}
\label{sec:background}

Social media has emerged as a dominant platform for sharing information online, with a global user base exceeding 4.59 billion in 2022, expected to approach six billion by 2027 \cite{Bakshy.2015,Pew.2021}. The shift from traditional media to social media has essentially transferred the responsibility of content quality control from trained journalists to everyday users \cite{Kim.2019} and created a fertile environment for the proliferation of misinformation \cite{Shao.2016}. Numerous studies have examined the spread of misinformation on social media, suggesting that false information spreads more viral than the truth \cite{Friggeri.2014,Vosoughi.2018,Solovev.2022b,Proellochs.2021b,Proellochs.2021a,Proellochs.2023}. Viral misinformation on social media can have severe real-world consequences, posing risks not only to individuals but also to society as a whole \cite{Allcott.2017,Bakshy.2015,DelVicario.2016,Oh.2013,Bar.2023,Feuerriegel.2023}.

Containing the spread of misinformation on social media necessitates accurate identification approaches \cite{Pennycook.2019}. Current measures of identifying misinformation fall under two primary categories. The first entails human-based approaches that rely on professionals or fact-checking organizations like Politifact and Snopes to verify the veracity of posts \cite{Hassan.2017,Shao.2016}. The second category entails machine learning-based systems, which attempt to automatically classify misinformation by leveraging content-based elements (\eg, images, text, video), context-based elements (\eg, time and location), or propagation patterns \cite{Ma.2016,Qazvinian.2011}. However, both approaches exhibit inherent drawbacks. Verification performed by experts typically delivers reliable results but grapples with scalability owing to the scarcity of professional fact-checkers. Conversely, while detection powered by machine learning provides scalability, it frequently underperforms in terms of prediction accuracy \cite{Wu.2019}. Consequently, this indicates the necessity for approaches to fact-checking that combine accuracy with scalability.

As an alternative, recent research has suggested delegating the task of fact-checking misinformation on social media to non-experts in the crowd \cite{Micallef.2020,Bhuiyan.2020,Pennycook.2019,Epstein.2020,Ma.2023,Allen.2020,Allen.2021,Godel.2021}. The intuition is to harness the wisdom of crowds to identify misleading posts \cite{Woolley.2010}. Different from expert-based approaches, which are hindered by the limited pool of professional fact-checkers, community-based approaches make it possible to identify misinformation at a high volume \cite{Pennycook.2019}. Additionally, community-based fact-checking tackles the problem of user skepticism towards professional fact-checks \cite{Poynter.2019}. Existing works imply that although assessments from single users might be inconsistent and unreliable, they tend to be highly accurate when collated \cite{Woolley.2010}. Experimental research has shown that crowds can be remarkably accurate in recognizing misleading content on social media platforms, indicating that even fairly small ensembles of non-experts can achieve results comparable to those of experts \cite{Bhuiyan.2020,Epstein.2020,Pennycook.2019}.

Although the crowd might be \emph{capable} of correctly identifying misinformation, it does not automatically entail that all users will \emph{decide} to do so \cite{Epstein.2020}. Critical challenges encompass lack of engagement in critical thinking \cite{Pennycook.2019b}, politically motivated reasoning \cite{Kahan.2017}, and manipulation attempts \cite{Luca.2016}. Each of these behaviors can reduce the effectiveness of community-based fact-checking systems. For example, users might deliberately sabotage the fact-checking mechanism by reporting social media content that refutes their personal belief, irrespective of its actual truthfulness \cite{Luca.2016}. Furthermore, the stark polarization among social media users \cite{Conover.2011,Barbera.2015} might results in different interpretations of facts or even completely different sets of acknowledged facts \cite{Otala.2021}. Indeed, prior (small-scale) attempts towards community-based fact-checking, like TruthSquad, Factcheck.EU, and WikiTribune \cite{Oriordan.2019, Florin.2010} were confronted with quality issues regarding user-created fact-checks \cite{Bhuiyan.2020,Bakabar.2018}. This highlights the difficulty of implementing real-world community-based fact-checking systems that preserve both high level of quality and scalability. Core requirements to counter the aforementioned challenges encompass advanced rating systems and fact-checking guidelines that foster helpful context \cite{Epstein.2020,Godel.2021}.

In an attempt to address these challenges, the social media platform X (formerly Twitter) launched its community-based fact-checking system Community Notes (formerly known as ``Birdwatch'') \cite{Twitter.2021,Proellochs.2022a}. Different from earlier crowd-based fact-checking initiatives, Community Notes allows users to identify misinformation \emph{directly} on the platform. Community Notes also implements a rating mechanism that allows users to rate the helpfulness of other users' fact-checks. However, given the novelty of the platform, research studying how users interact with Community Notes is still relatively scant. Early works have primarily analyzed the targets of community fact-checkers \cite{Pilarski.2023,Saeed.2022,Allen.2022,Proellochs.2022a} and the spread of community fact-checked posts on X \cite{Drolsbach.2023a,Chuai.2023,Drolsbach.2023b,Chuai.2024a}. While politically motivated reasoning might pose challenges \cite{Allen.2022,Proellochs.2022a}, research suggests that community notes can successfully reduce users' belief in false content and their intentions to share misleading posts \cite{Chuai.2024a,Drolsbach.2024b}. Our study adds by studying the link between external sources in community-created fact-checks and their helpfulness.

\section*{Research Questions}
\label{sec:hypotheses}

\subsection*{Helpfulness of External Sources (RQ1)}

Community-based fact-checking systems can provide fact-checkers with the option to link to external sources (\ie, websites) to support their assessments of social media posts \cite{Twitter.2021,Proellochs.2022a}. Multiple considerations lead us to expect that fact-checks that make use of this option and do link to external sources are perceived as more helpful by other users. First, the presence of links to external sources is likely to make the fact-check more credible. Arguments tend to be more credible if they provide more information in support of the advocated position \cite{Schwenk.1986}. It is well documented that the advisees' perception of the credibility of the advisor is an important determinant of helpfulness \cite{Connors.2011,Li.2013}. Second, users may be unmotivated to invest the necessary effort of validating the assertions made in the fact-checks. In this scenario, more justifications for a position may make users more confident in their assessment \cite{Tversky.1974}. Third, the presence of links to external sources may reflect the fact-checkers's involvement and knowledge. The more effort and expertise the fact-checker puts into writing the fact-check, the more likely it is that it will provide high-quality information that presents helpful context to other users. Taking these arguments together, community fact-checks that link to external sources may contain more credible arguments presented by better-informed fact-checkers that are more helpful to other users. RQ1 states:

\vspace{0.3cm}
\emph{\textbf{RQ1:}} \emph{Are community fact-checks linking to external sources perceived as more helpful?}

\subsection*{Political Bias in External Sources (RQ2)}

The internet has given rise to an unprecedented prominence and popularity of politically biased sources of information \cite{Stroud.2010}. This raises the question of whether the effect of external sources in community-created fact-checks on helpfulness varies depending on their level of political bias. Fact-checkers can link to websites with high (\eg, partisan websites such as \url{breitbart.com}) or low political bias (\eg, mainstream media outlets). In general, politically biased sources tend to be perceived as less credible than non-biased sources \cite{Pornpitakpan.2004}. We expect that users perceive politically biased sources as less helpful because individuals have a well-developed association between credible sources and truthful information \cite{Fragale.2004,Traberg.2022}. In other words, it may be easy for individuals to rely on the simple-decision rule ``experts are usually correct'' when judging the likely authenticity of a fact-check. Furthermore, people are generally more persuaded by high-credibility sources \cite{Fragale.2004,Hovland.1951,Traberg.2022}, which can even make them more likely to agree with counter-attitudinal viewpoints \cite{Hovland.1951}. It is thus plausible that fact-checks that leverage source credibility are more likely to be helpful. Based on this reasoning, we hypothesize that fact-checks linking to external sources with high political bias are perceived as less helpful than those linking to external sources with low political bias. RQ2 states:

\vspace{0.3cm}
\emph{\textbf{RQ2.}} \emph{Is linking to low bias sources in community fact-checks is perceived as more helpful than linking to high bias sources?}

\subsection*{Partisan Asymmetry (RQ3)}

Politically biased sources typically have a distinct partisan leaning, favoring either conservative (\ie, right-leaning) or liberal (\ie, left-leaning) opinions \cite{Chuai.2025a}. At the same time, social media is characterized by ``us versus them'' mentality (\ie, a partisan-laden perception), which can result in the dismissal of viewpoints and facts from the political out-group \cite{vanBavel.2018}. Contextualized to community-based fact-checking, linking to politically biased sources may implicitly reveal information about the political orientation of the fact-checker -- which may be polarizing to users with opposing (political) views. Assuming a high level of political diversity among the users participating in community-based fact-checking (\ie, both fact-checkers and raters), this would imply that biased sources of either political side are less likely to be perceived as helpful by a large share of users. However, the assumption of high political diversity may not hold true in real-world community-based fact-checking systems such as X's Community Notes. The reason is that users engaging in community-based fact-checking are \emph{self-selected} and, thus, are not necessarily representative of the overall user base on social media or society as a whole. There is ample evidence that the political left and the political right use social media in different ways, a phenomenon known as \emph{ideological asymmetry} \cite{Gonzalez.2022}. For example, adherents of the left have been found to be less tolerable to the spread of misinformation and have greater trust in fact-checking \cite{Gonzalez.2022,Shin.2017}. It is thus conceivable that the self-selected fact-checking community is more likely to identify with one side of the political spectrum -- and that it may read its own political leanings into fact-checks. However, an understanding of whether politically biased sources are more helpful if they are left-leaning or right-leaning is missing. Hence, RQ3 states:

\vspace{0.3cm}
\emph{\textbf{RQ3:}} \emph{Are politically biased sources perceived as more helpful if they are left-leaning or right-leaning?}

\section*{Data and Empirical Model}
\label{sec:dataset}

\subsection*{Data}

\begin{figure}
	\centering
	\fbox{\includegraphics[width=.475\textwidth]{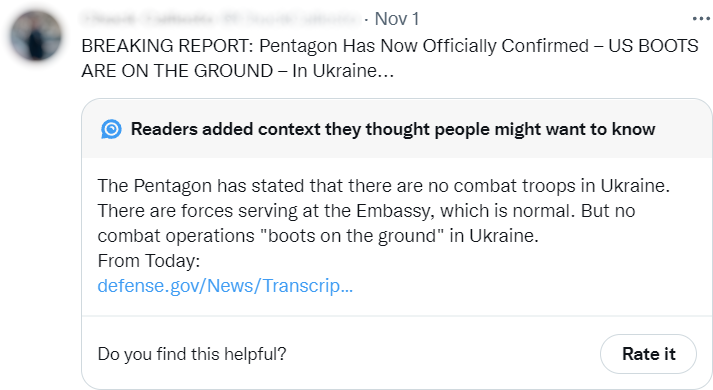}}
	\caption{\centering Example of a Community-Created Fact-Check (``Community Note'') on X.}
	\label{fig:birdwatch_example}
\end{figure}

This work examines the helpfulness of community-created fact-checks from X's Community Notes \cite{Twitter.2021}. Launched to the public in October, 2021, Community Notes is a novel platform to counter misinformation circulating on X through the power of collective intelligence. The Community Notes platform allows X users to identify posts they perceive as misleading and supplement them with \emph{textual} written notes, as illustrated in Fig.\ \ref{fig:birdwatch_example}. Community Notes are limited to 280 characters where each URL (\ie, website) accounts for a single character. Community notes can be attached to \emph{any} post on X. Following its submission, the note becomes accessible to other platform users. Community Notes also comes with a rating system allowing users to assess the helpfulness of notes submitted by others. Similar to other popular websites like \url{Amazon.com}, these user-generated ratings aim to identify and elevate the visibility of the most helpful and relevant context.

\subsubsection*{Data collection}
We retrieved \emph{all} Community Notes and corresponding original posts from the official roll-out of the Community Notes in October 2022 until June 2023 from the Community Notes site (\url{birdwatch.twitter.com}). Following earlier work on helpfulness \cite{Lutz.2022,Korfiatis.2012,Mudambi.2010}, we only consider fact-checks for which the helpfulness has been assessed at least once by users (\ie, fact-checks that received at least one helpful or unhelpful vote). The resulting dataset contains a total number of \num{41128} Community Notes (\ie, community-created fact-checks), and \num{2848825} ratings (\ie, helpfulness votes). We utilized the historical API provided by X to correlate the \emph{postID} referenced in every Community Note with the original post (\ie, the post that was subject to fact-checking) and collected the following information about each original post and the account of its author: (i) the number of followers, (ii) the number of followees, (iii) the account age, (iv) whether the user has been verified by X, (v) the post age.

\subsubsection*{Links to external sources}
Subsequently, we extracted each link to external websites from the text explanations in the Community Notes (see Fig.\ \ref{fig:birdwatch_example}). To this end, we implemented string extraction of links in the Python programming language with the help of the built-in \texttt{re} package. Each of the extracted links was then reduced to its {domain name} (\eg, \url{cnn.com}). Among all Community Notes, \SI{88.66}{\percent} contain at least one link to an external source. The majority of Community Notes with external sources contained only a single link, whereas \SI{35.58}{\percent} of the Notes contained multiple external links. The most common sources in Community Notes are links to \emph{Media Outlets} and \emph{Public Authorities}, which represent \SI{50.09}{\percent} and \SI{18.25}{\percent} of all links in our dataset, respectively. This is followed by \emph{Social Media Posts} (\SI{13.97}{\percent}), \emph{Scientific Literature} (\SI{7.04}{\percent}), \emph{Encyclopedias} (\SI{5.82}{\percent}), and \emph{Third-Party Fact Checkers} (\SI{3.34}{\percent}). \SI{1.50}{\percent} of all links refer to \emph{Other} sources.

\subsubsection*{Political bias}
To determine the political slant of the external sources, we utilized the website Media Bias/Fact Check (\url{mediabiasfactcheck.com}), which provides assessments of political bias (left and right) for a great deal of websites. The bias ratings from Media Bias/Fact Check are a common choice in previous literature \cite{Cinelli.2021} and are based on criteria such as the factuality of reporting, one-sidedness, and strength of political affiliations. We used Media Bias/Fact Check to collect information about (i) the bias magnitude (low, medium, high), and (ii) the bias direction (left, undirected, right) of the external sources in Community Notes.

\begin{figure}[H]
	\centering
	\includegraphics[width=.475\textwidth]{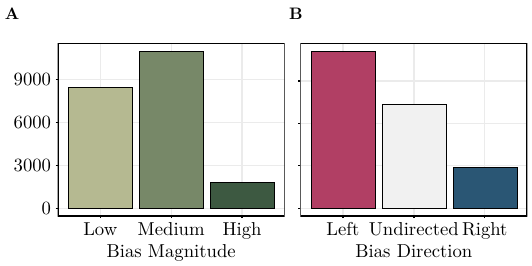}
	\caption{Distribution of political biases in Community Notes. (A) Bias magnitude ordered from Low to High. (B) Bias direction, separated into Left, Undirected, and Right.}
	\label{fig:bias_descriptives}
\end{figure}

By matching the bias rating from Media Bias/Fact Check to the extracted domain names, we were able to obtain bias scores for \SI{50.35}{\percent} of all links in Community Notes. Figure \ref{fig:bias_descriptives} illustrates the distribution of the detected bias magnitudes and directions. External sources with medium bias are most common in our sample (\SI{51.51}{\percent}), followed by low bias (\SI{39.82}{\percent}). Notes containing highly biased sources are relatively rare (\SI{8.67}{\percent}). Regarding the bias direction, left-leaning external sources are more prevalent in Community Notes with approximately \SI{52}{\percent} (\num{11089}) of Notes having a clear left-leaning bias, while only \SI{13.61}{\percent} of Notes have a clear right-leaning bias. Approximately \SI{34.35}{\percent} of external sources are politically neutral (i.e., undirected bias). Examples of the most common domains referenced in Community Notes are reported in Table \ref{tbl:domains}.

\begin{table}
	\centering
	{
		\footnotesize
		\begin{tabularx}{0.5\textwidth}{@{\hspace{\tabcolsep}\extracolsep{\fill}}llr }
			\toprule
			 & {\textbf{Domain}}        & {\textbf{Frequency}} \\
			\midrule
			\underline{Overall}                                \\
			 & {twitter.com}            & 5747                 \\
			 & {wikipedia.org}          & 2298                 \\
			 & {apnews.com}             & 1220                 \\
			 & {snopes.com}             & 1127                 \\
			 & {youtube.com}            & 1086                 \\
			\addlinespace
			\underline{Low Bias}                               \\
			 & {wikipedia.org}          & 2298                 \\
			 & {reuters.com}            & 979                  \\
			 & {cdc.gov}                & 715                  \\
			 & {nature.com}             & 524                  \\
			 & {thehill.com}            & 340                  \\
			\underline{Medium Bias}                            \\
			 & {apnews.com}             & 1220                 \\
			 & {snopes.com}             & 1127                 \\
			 & {nytimes.com}            & 835                  \\
			 & {npr.org}                & 819                  \\
			 & {washingtonpost.com}     & 765                  \\
			\underline{High Bias}                              \\
			 & {dailymail.co.uk}        & 295                  \\
			 & {state.com}              & 179                  \\
			 & {giszodo.com}            & 97                   \\
			 & {vox.com}                & 97                   \\
			 & {washingtonexaminer.com} & 91                   \\
			\bottomrule
		\end{tabularx}
		\caption{Most frequent domains referenced in the text explanations of Community Notes.}
		\label{tbl:domains}
	}
\end{table}

\subsection*{Variable Definitions}

We are interested in analyzing factors that determine the helpfulness of community-created fact-checks. To this end, the dependent variable is the number of $\textit{HVotes}$ (helpful votes), which denotes the number of users who voted ``Yes'' in response to the question ``Is this note helpful?'' The total number of users who responded to this question is denoted by $\textit{Votes}$.

The explanatory variables in our study can be divided into two groups: (1) variables that are given by the community-created fact-check (\ie, Community Note); and (2) variables that provide information about the original post (summary statistics and cross correlations are provided in the Supplementary Table S1 and Supplementary Figure S1).

\subsubsection*{Fact-checking variables}
Our key explanatory variable is \textit{External Source}, which is a binary label denoting whether a link to an external website has been provided as part of a Community Note ($=1$ if true, otherwise 0). Additionally, Media Bias/Fact Check provides ratings on the political bias of the external sources. We gather information about the magnitude and direction of the political biases for each Community Note. The resulting variable $\textit{Bias Magnitude}$ ranges from $0$ to $2$. Here a value of $0$ refers to sources with low bias, a value of $1$ refers to sources with medium bias, and a value of $2$ refers to high biased sources (in either political direction). If a Community Note contains multiple external links, we take the mean of the bias scores of the individual links. We follow the same approach to calculate individual scores for the $\textit{Bias Direction}$ ($-1$ for left-leaning, $0$ for undirected, and $+1$ for right-leaning.) For instance, if there are two links in a Community note pointing in different directions, the mean $\textit{Bias Direction}$ would be zero.

We use additional control variables to account for common content characteristics of the Community Notes that may affect their helpfulness: (i)~we control for the length ($\textit{Word Count}$) of the Community Notes (excluding links), (ii)~we calculate the $\textit{Text Complexity}$ using the Gunning-Fog readability index, and (iii)~we use the \texttt{sentimentr} package \cite{Rinker.2019} in combination with the built-in NRC lexicon \cite{Mohammad.2013} to measure the positive/negative $\textit{Sentiment}$ \cite{Feuerriegel.2025} of the Community Note.

\subsubsection*{Original post variables}
In our empirical analysis, we also control for characteristics of the original (\ie, the fact-checked) post. First, we control for the sentiment of the post ($\textit{Post Sentiment}$), analogous to the sentiment of the Community Note. Second, we control for the social influence of the author of the original post. The variables include the number of followers ($\textit{Followers}$), the number of followees ($\textit{Followees}$), the account age ($\textit{Account Age}$), whether the account has been verified by X ($=1$ if true, otherwise 0), and how many days have passed since the post was first published ($\textit{Post Age}$). Third, we use a binary variable $\textit{Political}$ denoting whether the original post covers a political topic ($=1$ if true, otherwise 0). To this end, we fine-tuned (and manually validated) the pre-trained TwHIN-BERT language model \cite{Zhang.2022} for our task (see SI, Supplement D for implementation details). The classifier achieved a high macro-averaged $F_{1}$ score of 0.755 in predicting topic labels.

\subsection*{Model Specification}

Following previous research modeling helpfulness \cite{Yin.2016,Lutz.2022}, we model the number of helpful votes, $HVotes$, as a binomial variable with probability parameter $\theta$ and $Votes$ trials. Our key explanatory variable that allows us to analyze \emph{RQ1} is $\textit{External Source}$, a binary variable denoting whether a Community Note includes links to external sources ($=1$ when true, 0 otherwise). We control for multiple content characteristics of Community Notes, namely, the length ($\textit{Word Count}$), text complexity ($\textit{Text Complexity}$), and $\textit{Sentiment}$. Furthermore, we control for various characteristics of the fact-checked post. The control variables include the number of $\textit{Followers}$ and $\textit{Followees}$, the account age ($\textit{Account Age}$), whether the account is $\textit{Verified}$, the post age ($\textit{PostAge}$), the post sentiment ($\textit{Post Sentiment}$), and a binary dummy indicating whether the fact-checked post covers a political topic ($=1$ when true, 0 otherwise). This yields the following regression model:
\vspace{-.7\baselineskip}

{\small
	\begin{flalign}
		\logit  (\theta) & = \,\beta_0 + \beta_{1} \,\textit{External Source} + \,\beta_{2} \,\textit{Word Count} + \beta_{3} \,\textit{Text Complexity}+ \beta_{4} \,\textit{Sentiment}       \label{eq:theta}                            \\ \nonumber
		                 & + \,\beta_{5} \,\textit{Followers} + \beta_{6} \,\textit{Followees} + \beta_{7} \,\textit{Verified} + \,\beta_{8} \,\textit{Account Age} + \beta_{9} \,\textit{Post Age} + \beta_{10} \,\textit{Post Sentiment} \\ \nonumber
		                 & + \,\beta_{11} \,\textit{Political} + \, \varepsilon , \nonumber
	\end{flalign}
	\begin{flalign}
		\hspace{1.6em} HVotes \sim Binomial[Votes, \theta],  \label{eq:binom} \hskip0.2\textwidth
	\end{flalign}
}
\normalsize
with intercept $\beta_0$ and error term $\varepsilon$. We estimate Eq.~\ref{eq:theta} and Eq.~\ref{eq:binom} using maximum likelihood estimation and generalized linear models. To facilitate the interpretability of our findings, we $z$-standardize all continuous variables, allowing us to compare the effects of regression coefficients on the dependent variable measured in terms of standard deviations.

In order to analyze RQ2 and RQ3, we focus on the subset of community-created fact-checks that contain at least one link to external sources rated by Media Bias/Fact Check. The key explanatory variable that allows us to analyze RQ2 is $\textit{Bias Magnitude}$, \ie, the severity of political bias of the links provided in Community Notes. To study RQ3, we include additional interaction term between $\textit{Bias Magnitude}$ and $\textit{Bias Direction}$, which allows us to analyze whether politically biased sources are more/less helpful depending on whether they are left-leaning or right-leaning. All controls are analogous to the previous model.

Note that we analyze a wide range of additional model variants as part of an extensive set of robustness checks. In all of these analyses, we observe consistent results.

\section*{Empirical Results}
\label{sec:results}

\subsection*{Helpfulness of External Sources (RQ1)}

We now analyze factors that determine the helpfulness of community-created fact-checks (RQ1). For this purpose, we draw upon a binomial regression model with the share of helpful votes as the dependent variable. The coefficient estimates for our primary explanatory variables are visualized in Fig.\ \ref{fig:coefs_helpfulness} (see Supplementary Table S2 for full estimation results).

Our findings suggest that the content characteristics of Community Notes play an important role in determining their helpfulness: the coefficients for \emph{Word Count} (coef.\ = $0.048$, OR = 1.049, $p<0.001$), \emph{Text Complexity} (coef.\ = $-0.033$, OR = 0.968, $p<0.001$), and \emph{Sentiment} (coef.\ = $0.018$, OR = 1.018, $p<0.001$) are positive and statistically significant. For a one standard deviation increase in the explanatory variable, the estimated odds of a helpful vote increase by $e^{0.048}-1 \approx$ \SI{4.92}{\percent} for \emph{Word Count}, \SI{3.25}{\percent} for \emph{Text Complexity}, and \SI{1.82}{\percent} for \emph{Sentiment}. Consequently, the perceived helpfulness of Community Notes is higher if they incorporate a more positive sentiment, are of greater length, and utilize less complex language.

We further note that the social influence attributed to the account that disseminates the original post has an effect on the perceived helpfulness of Community Notes. Here, the largest effect sizes are estimated for \emph{Verified}, \emph{Followers}, and \emph{Account Age}. The odds of receiving a helpful vote for Community Notes reporting posts from verified accounts are \SI{9.86}{\percent} higher (coef.\ = $0.094$, OR = 1.099, $p<0.001$) than for unverified accounts. A one standard deviation increase in the number of followers decreases the odds of a helpful vote by \SI{22.89}{\percent} (coef.\ = $-0.260$, OR = 0.77, $p<0.001$). A one standard deviation increase in the time since the account was published is associated with a \SI{3.92}{\percent} decrease (coef.\ = $-0.040$, OR = 1.049, $p<0.001$) in the estimated odds of a helpful vote. In sum, there is a lower level of helpfulness for posts from high-follower and older accounts, and a higher level of helpfulness for Community Notes fact-checking posts from verified accounts. Furthermore, we find that fact-checks for posts covering a political topic are significantly less helpful (coef.\ = $-0.050$, OR = 0.951, $p<0.001$).

To analyze \emph{RQ1}, we assess how the presence of links to external sources in Community Notes is linked to their helpfulness. The coefficient estimate for \emph{External Source} is \num{0.994} (OR = 2.70, $p<0.001$), which implies that the odds of Community Notes linking to external sources to be perceived as helpful are 2.70 times higher than for those not containing links to external sources. Notably, this is, by far, the largest effect size across all variables in our model.

\begin{figure}
	\centering
	\includegraphics[width=.475\textwidth]{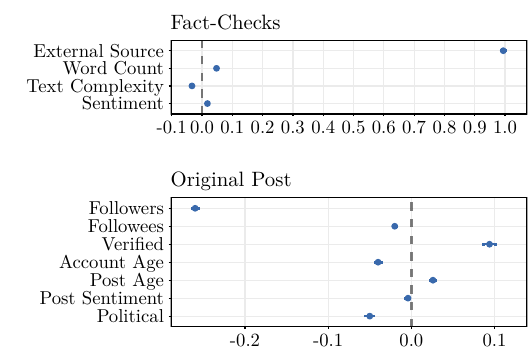}
	\caption{Binomial regression analyzing the helpfulness of external sources in explaining the share of helpful votes. Shown are coefficient estimates with \textbf{\sisetup{detect-all}\SI{95}{\percent}} CIs. Unit of analysis is the fact-check level \textbf{($N = 41,129$)}.}
	\label{fig:coefs_helpfulness}
\end{figure}

\subsection*{Political Bias in External Sources (RQ2)}

Next, we analyze the role of political bias regarding the helpfulness of external sources in community-created fact-checks (RQ2). For this purpose, we additionally include the variable $\textit{Bias Magnitude}$ into the regression model and restrict our analysis to Community Notes containing at least one link to an external source rated by Media Bias/Fact Check, resulting in \num{21307} observations. The control variables are analogous to the previous model.

\begin{figure}
	\centering
	\includegraphics[width=.475\textwidth]{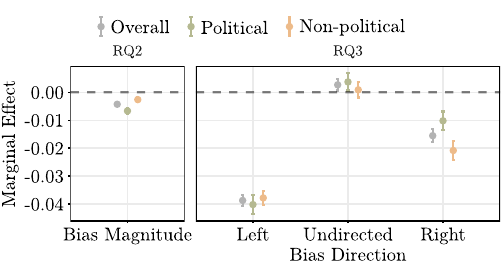}
	\caption{Marginal effects (with \textbf{\sisetup{detect-all}\SI{95}{\percent}} CIs) of bias magnitude (left panel) and bias direction (right panel) on the share of helpful votes. Unit of analysis is the fact-check level \textbf{($N = 21,307$)}.}
	\label{fig:coefs_subset}
\end{figure}

The coefficient estimates (see left panel in Fig.\ \ref{fig:coefs_subset} for marginal effects, and Supplementary Table S3 for full estimation results) imply that linking to politically biased sources in community-created fact-checks is perceived as significantly less helpful. Specifically, a one standard deviation increase in $\textit{Bias Magnitude}$ is associated with a \SI{2.08}{\percent} decrease (coef.\ = $-0.021$, OR = 0.979, $p<0.001$) in the odds of a Community Note being perceived as helpful. To put this number into perspective, this implies that a community-created fact-check providing a link to a highly biased website (\eg, Breitbart) is approximately \SI{7.18}{\percent} less likely to be perceived as helpful than a fact-check linking to a low biased website (\eg, Reuters).

\subsubsection*{Political vs.\ non-political posts}
We now assess the role of political bias for political vs.\ non-political posts. For this, we include an interaction term between $\textit{Bias Magnitude}$ and $\textit{Political}$ into our regression model. After including the interaction, the estimated coefficient for the direct effect of $\textit{Bias Magnitude}$ is still negative and statistically significant (coef.\ = $-0.013$, OR = 0.987, $p<0.001$). This implies that for non-political posts, a one standard deviation increase in $\textit{Source Bias}$ is associated with an \SI{1.29}{\percent} decrease in the odds of a Community Notes to be perceived as helpful. Furthermore, we observe that the coefficient of the interaction between $\textit{Source Bias}$ and $\textit{Political}$ is also negative and statistically significant (coef.\ = $-0.019$, OR = 0.981, $p<0.001$), which implies that the effect of bias in external sources is slightly less negative for political posts. For political posts, we can assess the effect size by calculating the exponent of the sum of the coefficients \cite{Buis.2010} of $\textit{Bias Magnitude}$ and $\textit{Bias Magnitude} \times \textit{Political}$. The resulting OR is \num{0.969}, which implies that a one standard deviation increase in $\textit{Bias Magnitude}$ reduces the odds of a vote being rated helpful by \SI{3.15}{\percent}. This implies that the (negative) effect of bias in external sources on helpfulness is stronger for political posts, compared to non-political posts.

\subsection*{Partisan Asymmetry (RQ3)}
Next we analyze whether there is a partisan asymmetry, \ie whether politically biased sources are more helpful if they are left-leaning or if they are right-leaning (RQ3). The marginal effects are visualized in the right panel of Fig.\ \ref{fig:coefs_subset} (see Supplementary Table S4 for full estimation results).

We find that the presence of both right-leaning and left-leaning biased sources in community notes significantly reduces their perceived helpfulness. However, the magnitude of the effect sizes significantly differ: the inclusion of left-leaning biased sources reduces the perceived helpfulness by, on average, \SI{3.80}{\percent} (ME $= -0.039$, OR = 0.962, $p < 0.001$), whereas the inclusion of right-leaning biased sources reduces the perceived helpfulness by, on average, \SI{1.54}{\percent}  (ME $= -0.016$, OR = 0.985, $p < 0.001$). In contrast, when sources with undirected political bias are included, the perceived community note's helpfulness increases by, on average, 0.26 percentage points (ME $= 0.003$, OR = 1.00, $p < 0.05$). Overall, this implies that external sources with the same level of political bias are rated as the least helpful if they are left-leaning.

\subsubsection*{Political vs.\ non-political posts}
We assess the role of bias direction for political vs.\ non-political posts. To this end, we extend our regression model with an additional interaction term between $\textit{Bias Direction}$ and $\textit{Political}$. When a right-leaning biased source is included in a community note concerning a political post, there is, on average, a \SI{-1.02}{\percent} decrease in perceived helpfulness (ME $= -0.010$, OR = 0.990, $p < 0.001$). However, the inclusion of a similar bias in a community note on a non-political post results in a larger, \SI{2.07}{\percent} decrease in perceived helpfulness (ME $= -0.021$, OR = 0.979, $p < 0.001$). This variation between the two effects is statistically significant ($p < 0.001$). For left-leaning biased sources and sources with undirected political bias, we observe no statistically significant differences between for political vs. non-political posts (each $p > 0.05$).

\subsection*{Exploratory Analyses \& Robustness Checks}
\label{sec:robustness}

Multiple exploratory analyses and checks validated our results and confirmed their robustness. Specifically, we (1) controlled for fact-checks notes that contain multiple external sources, (2) analyzed helpfulness across different types of media categories (\eg, media outlets, scientific literature), (3) explicitly controlled for the level of factual reporting of websites, and (4) conducted a variety of additional robustness checks. In all of these checks, we find consistent results and our hypotheses continue to be supported. In the following, we provide a summary of the main findings.

\subsubsection*{Multiple External Sources}
Our main analysis focuses on the presence of at least one external source in community-created fact-checks (\ie, a binary variable). However, \SI{34.64}{\percent} of Community Notes contain multiple external links. We explicitly control for the number of links authors provide as part of their fact-check (see Supplementary Table S5). The coefficient for $\textit{Number of External Sources}$ are slightly positive and statistically significant (coef.\ = $0.076$, OR = 1.08, $p < 0.001$), implying that including multiple links to external sources in community-created fact-checks increase helpfulness.

\subsubsection*{Analysis Across Media Types} We further explore how the helpfulness of external sources varies across different media types. For this purpose, two trained research assistants manually assigned media categories (\eg, media outlets, scientific literature) to each external source in our dataset (multiple selection possible). The most common sources in Community Notes are links to \emph{Media Outlets} and \emph{Public Authorities}, which represent \SI{50.09}{\percent} and \SI{18.25}{\percent} of all links in our dataset, respectively. This is followed by \emph{Social Media Posts} (\SI{13.97}{\percent}), \emph{Scientific Literature} (\SI{7.04}{\percent}), \emph{Encyclopedias} (\SI{5.82}{\percent}), and \emph{Third-Party Fact Checkers} (\SI{3.34}{\percent}). \SI{1.50}{\percent} of all links could not be assigned to these categories and were labeled as \emph{Other}.

\begin{figure}
	\centering
	\includegraphics[width=.475\textwidth]{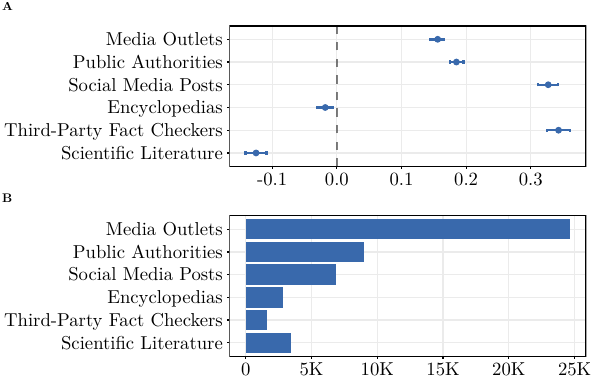}
	\caption{Analysis across media types. (A)~Results of a binomial regression analyzing the helpfulness of external sources varies across different media types. Shown are coefficient estimates with \textbf{\sisetup{detect-all}\SI{95}{\percent}} CIs. Unit of analysis is the fact-check level (\textbf{$N = 21,277$)}. (B)~Frequencies of media types across community notes.}
	\label{fig:coefs_robustness}
\end{figure}

Subsequently, we repeat our regression analysis with binary variables denoting the presence of the corresponding source categories as part of the Community Notes (see Supplementary Table S5). The coefficient estimates for media types and their frequencies are visualized in Fig.\ \ref{fig:coefs_robustness}. We find that community-created fact-checks linking to fact-checks from third-party fact-checkers (\eg, \url{snopes.com}) are perceived as particularly helpful (coef.\ = $0.343$, OR = 1.409, $p < 0.001$), followed by social media posts (coef.\ = $0.327$, OR = 1.387, $p<0.001$), and links to $\textit{Public Authorities}$ (coef.\ = $0.184$, OR = 1.202, $p<0.001$), and $\textit{Media Outlets}$ (coef.\ = $0.154$, OR = 1.166, $p<0.001$). In contrast, community-created fact-checks linking to $\textit{Scientific Literature}$ (coef.\ = $-0.126$, OR = 0.882, $p < 0.001$) and $\textit{Encyclepedias}$ (\eg, Wikipedia) (coef.\ = $-0.0019$, OR = 0.981, $p < 0.01$) are perceived as less helpful.

\subsubsection*{Additional Checks}
We performed an extensive series of supplementary analyses: (1)~we controlled for outliers in the dependent variables; (2)~we computed the variance inflation factors for each independent variable and validated that they are below the critical threshold of four; (3)~we included quadratic effects; (4)~we repeated our regression analysis by modeling the total count of votes (helpful and unhelpful) as the dependent variable; (5)~we re-coded bias magnitude and bias direction into a single variable (see Supplementary Table S6); (6)~we re-coded bias magnitude into a factor variable (see Supplementary Table S7). All the aforementioned analyses supported our findings.

\section*{Discussion}
\label{sec:discussion}

Our empirical findings contribute to research on misinformation on social media platforms and community-driven fact-checking. Whereas previous experimental studies have been primarily centered around the question of whether a crowd \emph{can} accurately evaluate content on social media \cite{Pennycook.2019}, an understanding of ``what makes community-created fact-checks helpful'' has remained largely absent. In this study, we hypothesized that linking to external sources is \emph{the} key determinant of helpfulness in community-based fact-checking (RQ1). Our primary rationale was that fact-checks are more credible and persuasive if they provide more information in support of their assertions. Furthermore, the presence of links to external sources may reflect the fact-checkers's expertise. Concordant with small-scale empirical analyses carried out during the pilot phase of community notes \cite{Proellochs.2022a}, we found strong support for this hypothesis: on average, the odds for Community Notes to be perceived as helpful were 2.70 times higher if they link to external sources. Notably, this effect size was larger than for any other considered predictor of helpfulness (\ie, content characteristics, author characteristics).

We further analyzed whether the link between external sources in community-created fact-checks and helpfulness varies depending on their level of political bias (RQ2). Our rationale was that community-created fact-checks leveraging on source credibility may be more likely to be effective. Consistent with this notion, we found that linking to high bias sources (\eg, ``alternative'' news outlets) in community-created fact-checks is perceived as less helpful. We also compared the helpfulness across various sub-categories of external sources. Here we found that community-created fact-checks linking to fact-checks from third-party fact-checking organizations (\eg, \url{snopes.com}) are perceived as particularly helpful. In contrast, community-created fact-checks linking to encyclopedias (\eg, Wikipedia) and scientific literature are perceived as less helpful.

Furthermore, our study provides new insights into the debate on whether political one-sidedness among the user base might hamper community-based fact-checking. The reason for these concerns is that users participating in community-based fact-checking may not be free of partisan motifs and political bias, but rather read their own political leanings into fact-checks. Hence, it is vital that there is a high level of political diversity among the users participating in community-based fact-checking. In this regard, our empirical findings are encouraging: although authors of community fact-checks are more likely to link to left-leaning sources, biased sources of either political side are rated as less helpful by other users (RQ3). This suggests that the rating mechanism on the community notes platform indeed penalizes one-sidedness and politically motivated reasoning.

From a practical perspective, social media platforms should closely monitor the potential of community-created fact-checking systems for three main reasons: (i)~they allow fact-checkers to identify misinformation at a large scale, (ii)~they address the trust problem with professional fact-checks, and (iii)~they identify misinformation that is of direct interest to actual social media users -- and which may go unnoticed by third-party fact-checking organizations. As such, our findings are of potential value for the design of more sophisticated community-based fact-checking systems to combat misinformation on social media. Specifically, our results suggest that ranking systems should put strong emphasis on links to unbiased external sources provided in fact-checks. Although helpful fact-checks can be identified through voting systems, accumulating high numbers of votes requires time. As a remedy, social media platforms may build on our our findings to develop systems that facilitate an early detection of potentially helpful fact-checks, thereby helping to prevent unhindered dissemination of misleading social media posts.

As with any other research, our study has a number of limitations. Although we performed an extensive series of robustness checks, there may be additional unobserved factors affecting users' perceived helpfulness of a specific fact-check that we cannot control for in our study. For instance, our approach struggles to account for subjective characteristics in the perception of raters (\eg, users' knowledge). 
Our study is also limited by the accuracy and availability of the bias ratings for websites, specifically those from Media Bias/Fact Check. However the bias ratings from Media Bias/Fact Check are a common choice in previous literature \cite{Cinelli.2021} and rely on distinctive source characteristics such as the factuality of reporting. Ultimately, our conclusions are confined to the sphere of community-based fact-checking on X's Community Notes platform. Further research is necessary to understand if the observed patterns are generalizable to other crowd-sourced fact-checking platforms.

\section*{Conclusion}
\label{sec:conclusion}

Community-based fact-checking systems require sophisticated rating systems and fact-checking guidelines that promote helpful context. In this work, we empirically investigate the helpfulness of the context provided in community-created fact-checks on X's community-based fact-checking system Community Notes. Our analysis suggests that linking to external sources is \emph{the} key determinant of helpfulness in community-based fact-checking. Furthermore, we find that the rating mechanism on the Community Notes platform successfully penalizes political one-sidedness in fact-checking. Our study has important implications for social media platforms that can utilize our results to optimize their fact-checking guidelines and promote helpful fact-checks.

\bibliography{literature}

\end{document}